\documentclass[11pt]{article}
\pdfoutput=1
\usepackage{amsmath,amsfonts,amssymb}
\usepackage{graphicx}

\newcommand{\be}{\begin{equation}}
\newcommand{\ee}{\end{equation}}
\newcommand{\ba}{\begin{eqnarray}}
\newcommand{\ea}{\end{eqnarray}}

\newcommand{\U}{\mathrm{U}}

\begin{document}
\newcommand{\todo}[1]{{\em \small {#1}}\marginpar{$\Longleftarrow$}}   
\newcommand{\labell}[1]{\label{#1}\qquad_{#1}} %{\label{#1}} %  

\begin{center} {\Large \bf Black Holes, Entanglement and Random Matrices}
\end{center} 
%\vskip 1cm   

\vspace{7mm}

Vijay Balasubramanian$^{a,b,}$\footnote{\tt email: vijay@physics.upenn.edu}, 
Micha Berkooz$^{c,}$\footnote{\tt email: micha.berkooz@weizmann.ac.il}, 
Simon F. Ross$^{d,}$\footnote{\tt email: S.F.Ross@durham.ac.uk}, 
Joan Sim\'on$^{e,}$\footnote{\tt email: j.simon@ed.ac.uk}
%\\

\vspace{5mm}

\bigskip\centerline{$^a$\it David Rittenhouse Laboratories, University of Pennsylvania,}
\smallskip\centerline{\it 209 S 33$^{\rm rd}$ Street, Philadelphia, PA 19104, USA}
%\bigskip\medskip
\bigskip\centerline{$^b$\it Initiative for the Theoretical Sciences, CUNY Graduate Center,}
\smallskip\centerline{\it 365 Fifth Avenue, New York, NY 10016, USA}
%\bigskip\medskip
\bigskip\centerline{$^c$\it Department of Particle Physics and Astrophysics, Weizmann Institute of Science,}
\smallskip\centerline{\it  Rehovot 76100, Israel}
%\bigskip\medskip
\bigskip\centerline{$^d$\it Centre for Particle Theory, Department of Mathematical Sciences,} \smallskip\centerline{\it Durham University, South Road, Durham DH1 3LE, UK}
%\bigskip\medskip
\bigskip\centerline{$^e$\it School of Mathematics and Maxwell Institute for Mathematical Sciences,}\smallskip\centerline{\it University of Edinburgh, King's Buildings, Edinburgh EH9 3JZ, UK}
\bigskip\medskip
%\vfil

\begin{abstract}
We provide evidence that strong quantum entanglement between Hilbert spaces does not generically create semiclassical wormholes between the corresponding geometric regions in the context of the AdS/CFT correspondence.  We propose a description of low-energy gravity probes as  random operators on the space of black hole states. We use this description to compute correlators between the entangled systems, and argue that a wormhole can only exist if correlations are large.  Conversely, we also argue that large correlations can exist in the manifest absence of a Lorentzian wormhole.  Thus the strength of the entanglement cannot generically diagnose spacetime connectedness, without information on the spectral properties of the probing operators. Our random matrix picture of probes also provides suggestive insights into the problem of ``seeing behind a horizon''.  
\end{abstract}

\section{Introduction}

Famously, the eternal black hole in AdS space, understood as a wormhole between two asymptotically AdS regions, is dual to two non-interacting conformal field theories in a thermally entangled state \cite{juan}. Generalizing this idea, it has been proposed that spacetime connectedness in AdS space is related to quantum entanglement in the dual field theory \cite{VanRaamsdonk:2010pw}, and, further, that entanglement is equivalent to the existence of wormholes in spacetime \cite{Maldacena:2013xja}.\footnote{In a related development based on the Ryu-Takayanagi expression for the entanglement entropy in field theory in terms of minimal surfaces in AdS space \cite{Ryu:2006bv}, the areas of bulk surfaces have been reconstructed from  a ``differential entropy'' measured from the entanglement structure of the field theory state \cite{Balasubramanian:2013lsa, Myers:2014jia}. This may be related to the proposal in \cite{Swingle:2012wq} to reconstruct the bulk spacetime fron the  entanglement structure of the field theory state using tensor network techniques from condensed matter physics.}  There is some controversy over how general the relation between entanglement and wormholes is. It was conjectured in \cite{Maldacena:2013xja} that entanglement should be {\it identified} with the existence of a wormhole (ER=EPR). However Marolf and Polchinski \cite{Marolf:2013dba} used the eigenvalue thermalization hypothesis (ETH) to argue that the local correlations in a typical entangled state are weak, and hence should not correspond to a semiclassical wormhole in the bulk. Shenker and Stanford \cite{Shenker:2013pqa,Shenker:2013yza}  found examples of special states corresponding to  long semiclassical wormholes, where the local correlations are weak but a smooth wormhole exists. In this paper we will examine this question using a model based on describing low-energy probes in the bulk as random matrices acting on the space of states of a black hole. In this random matrix model we will find a suppression of correlations in the typical state (unlike for the thermofield double state), in agreement with \cite{Marolf:2013dba}, and argue that this implies that these typical states do not have a semiclassical wormhole interpretation.  

We consider a Hilbert space ${\cal H} = {\cal H}_L \otimes {\cal H}_R$, where $\mathcal H_{L,R}$ are identical and dynamically independent factors.  
A particular entangled state in this Hilbert space is the thermofield double state
 \begin{equation}
 |\psi_\beta\rangle = \frac{1}{Z(\beta)} \sum_i e^{-\beta E_i/2} |i\rangle_L \otimes |i\rangle_R ,
 \label{thermalent}
 \end{equation}
where $E_i$ is the energy of the eigenstate labeled by $i$ in the $L$ and $R$ Hilbert spaces, and  $Z(\beta)$ normalizes the state. Tracing over $\mathcal H_L$ gives a thermal density matrix in $\mathcal H_R$.   This state can be thought of as a purification of the thermal density matrix and is identified in AdS/CFT with the eternal black hole in the bulk.  One Hilbert space factor is associated to each of the two asymptotic boundaries, and the entropy of the reduced density matrix on $\mathcal H_R$ is identified with the area of the horizon, that is, with the minimal cross-sectional area of the Einstein-Rosen bridge (wormhole) between the asymptotic regions.  Thus, the entropy of the reduced density matrix diagnoses the size of the wormhole. Furthermore, the entanglement in (\ref{thermalent}) gives rise to finite ``two-sided''  correlation functions  $\langle \mathcal O_L \mathcal O_R \rangle$  between operators supported on $\mathcal H_{L,R}$ respectively.  In AdS space, this correlator is computed  in a suitable approximation from spacelike geodesics which link the two boundaries of  spacetime through the wormhole. 

Now consider some more general entangled state on ${\cal H}_L \otimes {\cal H}_R$ which reduces to the thermal density matrix when one factor is traced over.  Consider CFT operators  dual to supergravity fields for which a probe approximation in the bulk is appropriate (that is, where the effects of back-reaction of this operator insertion can be neglected; we will assume in particular that the insertion changes the energy by an amount $\Delta E \lesssim T$) and where the operator dimension $\Delta \gg 1$ so that the geodesic approximation \cite{Balasubramanian:1999zv} to bulk correlators is reliable.\footnote{We will also include objects like D-branes in this class. These two limits can be simultaneously realized by making the temperature $T$ or the typical energy of the entangled states sufficiently large.}  Then the existence of a wormhole would imply that the two-point function between insertions of this operator in the two entangled copies of the field theory, $\langle \mathcal O_L \mathcal O_R \rangle$, will be given by a geodesic passing through the wormhole, and hence should be of roughly the same order as the two-point function in a single copy of the CFT, $\langle \mathcal O_R \mathcal O_R \rangle$; we would not expect it to be suppressed by any factor of the dimension of the Hilbert space.  Thus,  if the two-point function $\langle \mathcal O_L \mathcal O_R \rangle$ is exponentially suppressed relative to $\langle \mathcal O_R \mathcal O_R \rangle$ by factors involving the dimension of the space in which the entanglement occurs (that is, the entropy of the reduced density matrix in one copy of the Hilbert space), we take this as evidence that the state does not correspond to a semi-classical wormhole. 

In Sec.~\ref{motiv}, we will argue that the low-energy gravity probes we're interested in can be approximated as random matrices acting on the space of states of a black hole, because of the inability of low-energy supergravity modes to probe typical states of black holes \cite{ Balasubramanian:2005kk,  Balasubramanian:2005qu,Lashkari:2014pna}.    This leads to our key methodological innovation: the matrix elements of a given operator are modelled by a matrix drawn at random from some suitable ensemble. Given that the operator $\mathcal O$ can be modelled by a random matrix, we can approximate the correlator $\langle \mathcal O_L \mathcal O_R \rangle$ by an average over the operator ensemble. This gives us a simple calculation to test the validity of the wormhole interpretation for a given state. 

Modelling the probe by a random matrix is a stronger statement than the usual idea that excitations of a thermodynamic system thermalize to local thermal equilibrium; in an ordinary system like a lump of coal, we have some spatial resolution for our probes, so the excitation does not act totally randomly in that it will initially excite some local subset of degrees of freedom. Probing black holes is supposed to be more difficult, in that once the excitation has fallen into the black hole it is no longer localized to a subset of the degrees of freedom but acts truly randomly. 

In section~\ref{micro}, we treat the gravity probes as matrices acting on the states with energy in some narrow range $(E-\Delta, E+ \Delta)$, defining a microcanonical Hilbert space $\mathcal H_E$, of dimension $d_E=e^{S(E)}$. We use a uniform random matrix ensemble on  $\mathcal H_E$ to model the operators.  This allows us to do computations in states
 \begin{equation}
 |\psi_{\text{U}} \rangle= {1 \over d_E} \sum_{i,j \in \mathcal H_E} U_{ij}  |i\rangle_L \otimes |j\rangle_R 
 \label{psiU}
 \end{equation}
where $U_{ij}$ is some unitary matrix.   For all of these states, tracing over one Hilbert space space factor gives the maximally mixed density matrix
 \begin{equation} \label{max}
 \rho_E = \frac{1}{d_E} \mathbb{I}_E. 
 \end{equation} 
with entanglement entropy $S(E)$.  One of these states is the microcanonical analogue of  (\ref{thermalent}): $|\psi\rangle =  (1/Z(\beta) ) \sum_{i} e^{-\beta E_i} |i\rangle_L \otimes |i\rangle_R \sim e^{-S(E)/2} \sum_{i}  |i\rangle_L \otimes |i\rangle_R$.  The question is  whether the ``two-sided'' two-point functions  $\langle \psi_\text{U} | \mathcal O_L \mathcal O_R |\psi_\text{U}\rangle$ could be interpreted in terms of a wormhole in AdS. We find that it is suppressed relative to \eqref{thermalent} by a factor of $1/d_E \sim e^{-S}$, where $S$ is the entropy of the reduced density matrix.  By contrast, if we pick $U$ to be the identity (i.e. the state is the microcanonical analogue of the thermally entangled state (\ref{thermalent})) this suppression vanishes. Thus, the two-sided correlators in generic states $|\psi_\text{U}\rangle$ with the same entanglement entropy as the thermal state do not have the structure expected to allow a dual description in terms of a classical wormhole.  

 The uniform random matrix ensemble of operators on $\mathcal H_E$ provides a basic approximation to the properties of low-energy supergravity modes, but it does not fully capture the physics of the correlation functions of these operators, notably their time dependence. In section \ref{can} we show that by introducing more general matrix ensembles where we allow transitions between states of different energies, we can reproduce the exponential decay of correlators determined by the quasi-normal modes in the AdS description, supporting our  proposal that physics of probes of complex gravitational states can be understood in terms of the dynamics of random matrices.  The ensembles introduced in Sec.~\ref{can}  also allow us to make closer contact with the work of \cite{Marolf:2013dba} which uses the eigenvalue thermalization hypothesis.  

In Sec.~\ref{lowT} we turn to an important subtlety: at low temperatures, below the Hawking-Page transition, the thermofield double state \eqref{thermalent} is no longer dual to a black hole, and is rather described in the bulk as two copies of thermal AdS.  (More accurately, the latter saddle point of the gravitational path integral dominates the over the black hole saddle point.)   We will see via analytic continuation from the Euclidean theory that the two-sided correlators in this state remain large despite the absence of a Lorentzian wormhole.   This emphasizes that these correlators can only provide a necessary, and not a sufficient, condition for the presence of a  wormhole.   As we discuss, our  picture of black hole probes remains consistent with this sort of transition if, below the transition, operators no longer act like random matrices.  This seems natural because this is now an ordinary weakly interacting gas, where the probe will excite the gas locally. 

Section 6 discusses further implications of a random matrix description of operators. In particular such a description gives correlators a robust analytic structure in the complex time plane irrespective of the state or ensemble which they probe. We speculate that this could lead to a universal ``behind the horizon" continuation for the observables described by our random matrix approximation.  We also note that random operators give rise to a Wick-like scheme of contractions, independent of the usual one in terms of creation and annihilation operators, and suggest that this  gives rise to a modified perturbative scheme at the horizon. 
 
\paragraph{Relation to other work:} In \cite{Marolf:2013dba}, Marolf and Polchinski invoked the eigenvalue thermalization hypothesis to consider the behaviour of the ``two-sided'' two-point functions $\langle \mathcal O_L \mathcal O_R \rangle$ in typical entangled states on ${\cal H}_L \otimes {\cal H}_R$. They found the same suppression by $e^{-S}$. The ideas underlying our calculation are  similar, but we approximate gravity operators as random matrices, rather than assuming eigenvalue thermalization.   One might regard our computations as an effective model reproducing the physics that would arise if eigenvalues thermalize.  The idea that the amount of quantum entanglement is not enough to guarantee the existence of semiclassical wormholes was also advocated in \cite{Avery:2013bea}.    These authors dispute the interpretation of the eternal BH as a thermally entangled state \cite{juan}, but our results support the latter idea.

 In  \cite{Shenker:2013yza}, Shenker and Stanford provided examples of states which have weak local corelations but which do have a smooth classical dual.  (Other work on deformations of the thermofield double state includes \cite{Mathur:2014dia}.) However, as noted in  \cite{Shenker:2013yza}, these states are not generic and it seems difficult to construct a smooth geometry with the correct properties to give the dual of a generic state with weak correlations. We therefore interpret the exponentially suppressed two-point functions that we find as a signal of the absence of a geometric connection in the dual. The wormholes discussed in  \cite{Shenker:2013yza} have a natural interpretation as states close to the thermofield double  (\ref{thermalent}) in a sense that could be made precise using the random matrix approach. 
 
For the thermofield double state \eqref{thermalent}, two-sided correlations are large at $t=0$ but decay as time increases. In \cite{susskind1,susskind2} it has been speculated that this decay is related to the increasing complexity of the state, with the length $\ell$ of the wormhole in units of the AdS scale $L$ being given by 
\begin{equation}
  \frac{\ell}{L} \sim \frac{\mathcal C}{S}, 
\end{equation}
where $\mathcal C$ is the computational complexity of the state and $S$ is the entanglement entropy. This would correspond to the two-sided correlations scaling as 
\begin{equation}
  \langle {\cal O}_L {\cal O}_R \rangle \sim e^{- \Delta \frac{\mathcal C}{S}}\,,
\end{equation}
where $\Delta$ is the conformal dimension of $\mathcal O$. Our results are qualitatively consistent with this picture, as the states $|\psi_\text{U}\rangle$ are more complex than \eqref{thermalent}, and have weaker correlations. However, if we apply this interpretation to our results (equivalently, those of \cite{Marolf:2013dba}), it would correspond to assigning a complexity $\mathcal C \sim S^2$ to our state $|\psi_\text{U} \rangle$, which is much smaller than the maximal complexity $\mathcal C_{max} \sim e^S$. It will be interesting to understand this relation in more detail. 

There is some tension between our random matrix picture and the idea that the black hole microstates can be modelled by smooth geometries or extended D-brane configurations \cite{Lunin:2001jy,Bena:2004de}, as one would expect probes of such local objects to excite them locally as with ordinary thermodynamic systems. The resolution may lie in unknown aspects of the dynamics of such microstate geometries.  One could also envision some splitting of the Hilbert space into subspaces such that the operator acts as random operators within each subspace.

\section{Random operators}
\label{motiv}

Our key idea is that operators corresponding to low-energy gravity probes act randomly on the space of states associated with a black hole. We first want to clarify the extent of our claim. Consider a typical pure CFT state $|\psi \rangle$ dual to a large black hole.\footnote{That is, where the black hole is the dominant contribution to the dual bulk description; this excludes low-temperature states where the bulk description is thermal AdS, see comments on this in section~\ref{lowT}.} We act on this state with some CFT operator $\mathcal O$ dual to a supergravity mode. There is considerable evidence that local operators in the CFT which do not significantly change the total energy are insensitive to the particular details of the state $|\psi \rangle$ \cite{Balasubramanian:2005kk,Balasubramanian:2005qu,Lashkari:2014pna}. Therefore we would expect the transitions between such states produced by the action of this operator to be essentially random.

However, such a random matrix model is not expected to capture all of the physics of the operator insertion. At short times, the change in the state $|\psi \rangle$ retains some structure which reflects which operator $\mathcal O$ we acted with. This is indicated for example by the two-point correlation function of an operator with its conjugate, which at very short distances is dominated by the singular part of the OPE. We refer to this as the structured part of the operator. But at longer times (presumably compared to some thermalization scale of the state), or when acting on the degrees of freedom associated with a black hole, the ability to know the final state after acting with $\mathcal O$ is exponentially small. We call this part of the operator's action the unstructured part of the operator, and it is this that we want to model by a random matrix. 

The distinction can be easily understood in the dual bulk perspective in the geodesic approximation. The structured part of a correlation function is associated with geodesics which remain far from the black hole horizon, while the unstructured part involves geodesics which pass close to the horizon, or disconnected paths where one particle is absorbed by the black hole and another one is emitted in the Hawking radiation. That this is a useful diagnostic can be seen by considering situations in which the operator will remain structured for a long time, for example by considering operators with large angular momentum on the global AdS black hole, and noting such cases are reflected in the existence of geodesics which stay outside the black hole for a long time. Thus, even though modeling the gravity operators as random operators does not capture all aspects of the operators, it should capture precisely the part that we need. This is because the issue of whether there is a semiclassical wormhole or not is an issue of what happens behind the horizon. 

Thus, {\it low energy gravity modes act, to a good approximation, as random matrices on the states of the black hole}. In other words, gravity modes encode the minimum possible information about the actual microstates of black holes. We will thus treat the matrix elements of a given operator $\mathcal O$ as drawn at random from a suitable matrix ensemble. We will define the ensembles of interest in the next two sections. 

In this paper, we use this random matrix description to provide a criterion for the existence of a wormhole in the dual gravitational description of a typical entangled state. As argued in the introduction, we would expect states with a wormhole interpretation to have the property that the two-sided correlator $\langle \mathcal O_L \mathcal O_R \rangle$ is of the same order as the one-sided correlator  $\langle \mathcal O_R \mathcal O_R \rangle$. Given the random matrix description, we can approximate these correlators by considering the  average over the matrix ensemble that the operators are drawn from. For simple matrix ensembles, it is then easy to test this criterion in generic entangled states. 

The use of an ensemble average is also supported by the fact that in the geodesic approximation, the bulk two-point function calculation is largely insensitive to the details of the individual operator being considered, depending only on its conformal dimension. One might however still be concerned that the average could be suppressed relative to the value for a particular operator by phase cancellation. But in section \ref{micro} we will see that when we consider the state \eqref{thermalent} the average remains of the same order as for a given operator, which provides some evidence against this possibility. We will also see that standard deviations in the ensemble averages are exponentially small. 

\section{Entanglement vs wormholes: fixed energy}
\label{micro}

In this section, we consider entangled states which involve energy eigenstates lying in a narrow range of energies, and restrict attention to operators acting within this energy range. That is, we work with states  belonging to the subspace ${\cal H}_E\otimes{\cal H}_E \subset {\cal H}_L\otimes {\cal H}_R$, where ${\cal H}_E$ contains exact energy eigenstates $|i\rangle \in {\cal H}_E$ with eigenvalues $E_i \in [E-\Delta,\,E+\Delta]$, and we assume that there is a large density of states at these energies.  Entangled states can be written in this energy basis as 
\begin{equation}
  |\psi_c \rangle = \sum_{i,j} c_{ij} |i,j\rangle \quad \text{with} \quad \sum_{i,j} |c_{ij}|^2 = 1
\end{equation}
where $|i,j\rangle = |i\rangle_L \otimes |j\rangle_R$.

A particularly interesting subset of quantum pure states in ${\cal H}_E \otimes {\cal H}_E$ is defined by the property that tracing over ${\cal H}_L$ gives rise to the microcanonical ensemble, i.e. the maximally mixed density matrix 
\begin{equation}
  \rho_E = \frac{1}{d_E}\sum_{i\in {\cal H}_E} |i\rangle\langle i| = \frac{\mathbb{I}_E}{d_E}\,,
 \label{eq:omegaE}
\end{equation}
where $d_E=e^{S(E)}$ is the dimension of ${\cal H}_E$. We will denote this set of states in the Hilbert space by ${\cal H}_U$.\footnote{Note that $\mathcal H_{\text{U}}$ is not a subspace of the Hilbert space as a vector space, as the requirement that the reduced density matrix is \eqref{eq:omegaE} is not a linear constraint on the Hilbert space.} It includes all the states in ${\cal H}_E\otimes {\cal H}_E$ of the form
\begin{equation}
  |\psi_{\text{U}} \rangle = e^{-S/2}\sum_{i,j} U_{ij} |i,j\rangle
\label{eq:gmicro}
\end{equation}
for any unitary matrix $\text{U} \in \U(d_E)$. These states have the same amount of entanglement as the state
\begin{equation}
|\psi_{\text{micro}}\rangle = |\psi_{\mathbb{I}}\rangle = e^{-S/2} \sum_{i\in{\cal H}_E} |i,i\rangle
\end{equation}
which provides the standard purification of the single sided microcanonical density matrix.\footnote{This is the microcanonical ensemble equivalent of the thermofield double state \eqref{thermalent}.} Restricting to states with the same reduced density matrix is useful because it allows us to see clearly that the details and not just the overall amount of entanglement between the two Hilbert spaces plays a key role in the emergence of a smooth wormhole. We will study how single sided and two sided correlators behave in various $|\psi_c\rangle$ and $|\psi_\text{U}\rangle$. 

In subsection 3.1 we define the operator ensemble we consider, which is just the ensemble of gaussian random matrices in  ${\cal H}_E$. In section 3.2 we evaluate single sided correlators in the various states. In section 3.3 we compute the two sided correlators for various states, and  compare them with the single sided correlators on the same states. In section 3.4 we compute the standard deviations of the various correlators.

\subsection{Operator averaging \& random matrices}
\label{sec:micro}

To model  operators that act within ${\cal H}_E$ we will assume that  operator matrix elements are drawn from the simplest Gaussian distribution 
\begin{equation}
  {\cal F}_{r}=\frac{1}{Z^M_{\text{r}}} dM_{ij}dM_{ij}^*\,e^{-\gamma \text{tr}\left(M\,M^\dagger\right)},
 \label{eq:microens}
\end{equation}
where we denote the  matrix elements $\langle i| {\cal O} | j\rangle$ as $M_{ij}$ $\forall\, |i\rangle,\, |j\rangle \in  {\cal H}_E$, and $Z_{r}^M$ is a normalization factor, chosen so that $\int {\cal F}_{\text{r}}=1$. We will refer to this as the {\it restricted} operator ensemble, as it applies to operators which are restricted to act within $\mathcal H_E$. This ensemble is assumed to be universal for all operators acting within this Hilbert space. We consider calculations where we take the ensemble average within the ensemble of operators \eqref{eq:microens}, keeping the state $| \psi_c \rangle$ fixed. 

This choice of ensemble is motivated by simplicity: it is the gaussian matrix ensemble invariant under unitary transformations of $\mathcal H_E$ which depends only on the dimensionality of ${\cal H}_E$.\footnote{This is true assuming that we do not impose any restrictions of hermiticity or unitarity on ${\cal O}$.} The gaussian assumption amounts to a sort of free-field approximation for the operators, as in the ensemble \eqref{eq:microens} the only non-trivial connected correlation function is the two point function
\begin{equation}
  {\bf E}\left(M^*_{ij}M_{kl} \right) =  \frac{1}{\gamma} \delta_{ik}\delta_{jl}\,.
 \label{eq:2ptmc}
\end{equation}
(We will use the notation ${\bf E}$ to stress that we are taking expectation values in our operator ensemble.) Thus, when we insert operators in higher-point correlation functions, the ensemble expectation values will be determined by a Wick-like pair-wise contraction of insertions of operators using \eqref{eq:2ptmc} (after appropriately summing over the indices). This should be related to the free field approximation in the bulk spacetime, which is valid at leading order in $N$ in the large $N$ limit of the CFT. We will comment further on corrections from including higher order polynomials in the ensemble measure and the perturbation theory around this free-field behaviour in section \ref{pert}. 

The ensemble involves a single parameter $\gamma$, which depends on the energy $E$ used to define ${\cal H}_E$. The scaling of $\gamma$ with $E$ can be determined by considering the  ``inclusive cross section'' - starting from a given initial state $|i_0\rangle \in {\cal H}_E$, acting on it by the operator ${\cal O}$, and ending in all possible states (in ${\cal H}_E$). This gives  
\begin{equation}
  {\bf E}\left(\sum_{k\in {\cal H}_E} |\langle k | {\cal O} | i_0\rangle|^2\right) = \frac{e^{S(E)}}{\gamma},
\label{eq:gfix}
\end{equation}
so requiring that the inclusive cross-section is finite implies 
\begin{equation}
\gamma = \hat \gamma e^{S(E)},
\end{equation}
that is, the parameter $\gamma$ should scale like the dimension $d_E= e^{S(E)}$ of $\mathcal H_E$.

\subsection{Single sided correlators}
\label{sec:singlemicro}

Since our criterion is based on a comparison between single sided and double sided two-point functions (to avoid issues related to operator normalization), we will begin by computing the operator ensemble average of single sided correlators using \eqref{eq:microens}. 

We will begin with $|\psi_\text{U}\rangle$. Since we are computing single sided correlators we can reduce to the single sided density matrix \eqref{eq:omegaE} first:
\begin{equation}
\begin{aligned}
\langle \psi_\text{U} | {\cal O}^\dagger_R(t){\cal O}_R(0) | \psi_\text{U} \rangle &= \text{tr}_{{\cal H}_R}\left(\rho_E\,{\cal O}^\dagger_R(t){\cal O}_R(0)\right) \\
  &= e^{-S} \sum_{i\in {\cal H}_E,n\in {\cal H}_R} e^{i(E_i-E_n)t} |\langle n |{\cal O}_R |i \rangle|^2
\end{aligned}
\end{equation}  
Under the assumption that ${\cal O}$ acts within ${\cal H}_E$ (which allows us to sum only over $|n\rangle\in{\cal H}_E$), we can easily estimate the size of this two-point function at $t=0$ to be of order one
\begin{equation}
  {\bf E}\bigl( \langle \psi_\text{U} | {\cal O}^\dagger_R(0){\cal O}_R(0) | \psi_\text{U}\rangle\bigr) = e^{-S}
  \sum_{i\in {\cal H}_E,n\in {\cal H}_E}  {\bf E} \bigl( |\langle n |{\cal O}_R |i \rangle|^2 \bigr) = e^{-S}\,\frac{e^{2S}}{\gamma} =\frac{1}{\hat\gamma}\,.
\label{eq:1sided}
\end{equation}
Our choice of scaling for $\gamma$ thus has the nice consequence that correlators allowing a semiclassical gravitational interpretation are order one, i.e. they do not scale in the dimension of the microcanonical ensemble $d_E=e^S$. Computing the full time dependent operator ensemble average two-point function is just as easy and it is given by
\begin{equation}
  {\bf E}\bigl(\langle \psi_\text{U} | {\cal O}^\dagger_R(t){\cal O}_R(0) | \psi_\text{U}\rangle\bigr) = \frac{e^{-S}}{\gamma} \sum_{i,n} e^{i(E_i-E_n)t} =
   \frac{e^{-2S}}{\hat\gamma} \bigl|  \text{tr}_{{\cal H}_E} (W(t)) \bigr|^2
\label{eq:sngla}
\end{equation}
where $W(t) = e^{-iHt}$. The time dependence is generated by the slight variations in the energy in the range from $E-\Delta$ to $E+\Delta$, which produces only a very slow time variation of the correlator, on the times scales where we resolve these small energy differences. We will have a detailed discussion of time dependence in section \ref{can}, where we consider an ensemble of operators which includes larger energy transitions, and we can model the bulk gravitational time dependence successfully. 

We could consider more general states $|\psi_c \rangle$ in a similar way, but one easily sees that the correlators will be qualitatively the same as the ones above for any state whose reduced density matrix is localised in ${\cal H}_E$. In fact they are qualitatively the same even for pure states $|\psi_\text{V} \rangle = \sum_i V_i |i\rangle  \in {\cal H}_E$,
\begin{equation}
  {\bf E} \bigl( \langle \psi_\text{V} | {\cal O}_R^\dagger(t) {\cal O}_R(0) | \psi_\text{V} \rangle \bigr) = \frac{e^{-S}}{\hat\gamma} \langle \psi_\text{V} | W(t)^* | \psi_\text{V} \rangle  \text{tr}_{{\cal H}_E}( W(t) ). 
\end{equation}
This is a straightforward consequence of our starting assumption that the operators act randomly on states in ${\cal H}_E$. If the operators have simple spectral properties - i.e., they are drawn randomly from a simple matrix ensemble  - then single sided correlators can't even distinguish whether the state is a density matrix or a pure state. For two sided correlators, the story will be different, but for a single sided observer, measuring any finite number of correlators will not distinguish between a pure state, the thermal state, or any other mixed state with the same macroscopic quantum numbers. 

\subsection{Two-sided correlators}

We can now compute two-sided correlators, i.e. those involving operators acting on both Hilbert spaces ${\cal H}_L$ and ${\cal H}_R$.  We first review the analytic continuation between the  one-sided and two-sided correlators in a thermal ensemble, which teaches us that we want to consider two-sided correlators where the operator on ${\cal H}_L$ is time reversed.\footnote{In general, the states in ${\cal H}_L$ are the CPT conjugates of the ones in ${\cal H}_R$. Since we are primarily interested in discussing the time dependence in our correlators, we stressed the time reversal.}  We then discuss the calculation of these two-sided correlators in the operator ensemble average, and compare to the one-sided case. 

\subsubsection{Mapping of operators}
\label{sec:map}

In the thermal ensemble, we can define operators acting on ${\cal H}_L$ in terms of the action on $\mathcal H_R$ by the formula (see for example \cite{raju})
\begin{multline}
\text{tr}\bigl( \rho_{\beta}  {\cal O}_{R,\alpha_1}(t_1).... {\cal O}_{R,\alpha_n}(t_n) {\cal O}_{R,\beta_l}({t'}_l+i\beta/2))... {\cal O}_{R,\beta_1}({t'}_1+i\beta/2)\bigr) =\\
=\langle \Psi | {\cal O}_{R,\alpha_1}(t_1)....{\cal O}_{R,\alpha_n}(t_n) {\tilde{\cal O}}_{L,\beta_1}({t'}_1)... {\tilde{\cal O}}_{L,\beta_l}({t'}_l)| \Psi \rangle
\end{multline}
Here $t_1, \cdots t_n$ and $t_1', \cdots t_n'$ are separately time ordered from earliest to latest times, and
  the $\alpha_i$ and $\beta_i$ simply label the operators.   Also, $\rho_\beta$ is the canonical thermal density matrix on ${\cal H}_R$, while 
\begin{equation}
|\Psi \rangle = \frac{1}{\sqrt{Z(\beta)}} \sum_\alpha e^{-\beta E_\alpha/2} | \alpha,\alpha\rangle
\end{equation}
is the thermofield double state.   For the particular case of two point functions,  inserting the thermal density matrix on the left hand side and the state $|\Psi\rangle$ on the right hand side gives, 
\begin{equation}
 \sum_{ij} e^{-\beta E_i} 
\langle  i | {\cal O}_{R,\alpha_1}(t) |j\rangle 
\langle j | {\cal O}_{R,\beta_1}({t'}+i\beta/2)) |i \rangle 
=\sum_{i,j} e^{-{\beta\over 2}(E_i+E_j)} 
\langle  i | {\cal O}_{R,\alpha_1}(t) |j\rangle 
\langle i | {\tilde{\cal O}}_{L,\beta_1}({t'})) |j \rangle
\end{equation}
where on the left hand side we inserted a complete set of states $|j\rangle$.    The two terms in the exponential on the right hand side come from the bra and the ket in the correlator.   Next, we can write the imaginary time translation of $O_{R,\beta_1}$ in the second term on the left hand side as $O_{R,\beta_1}(t'+i\beta/2) = e^{i(i\beta/2)H} O_{R,\beta_1}(t') e^{-i(i\beta/2)H}$.  Acting on the state vectors on either side of $O_{R,\beta_1}(t')$ this produces exponential factors that now match between the left and right hand sides.   So we can conclude that $\langle j | O_{R\beta_1}(t') |i\rangle = \langle i | \tilde{O}_{L,\beta_1}(t') |j \rangle$.   In other words $\tilde{O}_{L,\beta_1}(t') = O_{R,\beta_1}(t')^T$.   There is one final subtlety -- in the conventional thermofield double description and in the eternal black hole, global time is defined to run backwards in the second (L) copy.  To be consistent with this convention we should flip the direction of time for the $O_L$ operators    If we choose $t=0$, $t'=0$, to form the initial Cauchy surface and flip the time direction in ${\cal H}_L$ to align time with global time on both sides, we finally have
\begin{equation}
{\cal O}_L(t) = {\tilde{\cal O}}_L(-t)= {{\cal O}}_R(-t)^T\,.
\label{eq:opmap}
\end{equation}
In a general state $|\psi_c \rangle$, we will therefore consider correlators between an operator $\mathcal O_R$ acting on $\mathcal H_R$ and the operator $\mathcal O_L$ acting on $\mathcal H_L$ defined by \eqref{eq:opmap}. Comparing this two-sided correlator to the one-sided correlator  $\langle \mathcal O_R \mathcal O_R \rangle$ will give us our criterion for the existence of wormholes.  

\subsubsection{Two-sided correlators and semiclassical ER bridges}

Given a pure state $|\psi_\text{U}\rangle\in {\cal H}_L\otimes {\cal H}_R$, the two point two sided correlator is then
\begin{equation}
  \langle \psi_\text{U} | {\cal O}^\dagger_R(t){\cal O}_L(0)| \psi_\text{U} \rangle = e^{-S(E)}\sum_{i,j,k,l} U_{ij}\,U^\star_{kl}\,e^{i(E_l-E_j)t}\,\langle i|{\cal O}_R(0)|k\rangle\langle l|{\cal O}^\dagger_R(0)|j\rangle\,.
\end{equation}
The ensemble average is
\begin{equation}
\begin{aligned}
  {\bf E}\bigl( \langle \psi_\text{U} | {\cal O}^\dagger_R(t){\cal O}_L(0)| \psi_\text{U} \rangle \bigr) &=e^{-S(E)} \sum_{i,j,k,l} U_{ij}\,U^\star_{kl}\,e^{i(E_l-E_j)t}\, {\bf E}\bigl(M^\star_{jl}\,M_{ik} \bigr) \\
 &= \frac{e^{-2S(E)}}{\hat\gamma}|\text{tr}\left(U\,W(t)\right)|^2 ,
\label{eq:2ave}
\end{aligned}
\end{equation}
where we used \eqref{eq:2ptmc} and $W(t) = e^{-itH}$. 

We would like to compare this expression with the single sided two-point function  \eqref{eq:sngla}. 
It is enough to focus on $t=0$. Since the trace is at most of order $d_E = e^{S(E)}$, the two sided correlator is bounded by the single sided one \eqref{eq:1sided}, and they are the same only when $\text{U}\propto \mathbb{I}_E$. We interpret this as saying that the wormhole connecting the two spaces is as large or as semiclassical as it can be when we have the standard purification of the microcanonical density matrix.

For most choices of $\text{U}$, the correlator will be much smaller. To determine the value for a typical $\text{U}$, we can consider now drawing $\text{U}$ itself uniformly from the ensemble of random unitary matrices, corresponding to choosing a typical state in $\mathcal H_\text{U}$.  
%We can diagonalize $\text{U}=\text{diag}(e^{i\theta_1},....e^{i\theta_{d_E}})$ to write the equal time correlator as 
%\begin{equation} \label{2sided}
 %{\bf E}\bigl( \langle \psi_\text{U} | O_R^\dagger (0) O_L(0) |\psi_\text{U} \rangle\bigl)  =\frac{1}{\hat\gamma}\bigl|e^{-S} \sum_{j=1}^{d_E} e^{i\theta_j} \bigr|^2 .
%\end{equation}
%Since the phases $\theta_j$ parameterize the Cartan subalgebra of $\SU(d_E)$ and the measure is uniform on them, we are just adding $d_E$ random complex numbers of unit size.  That is, it is the sum of $d_E$ vectors  of unit length and random orientation in a plane. Their sum gives rise to a drunkard's walk, so  $\text{tr}\,\text{U}  = \sum_j e^{i \theta_j} \sim \sqrt{d_E}$, and 
\begin{equation}
  {\bf E}_\text{U} \bigl ( {\bf E} \bigl( \langle  \psi_\text{U} | {\cal O}^\dagger_R(0){\cal O}_L(0)| \psi_\text{U} \rangle \bigr)\bigr)= \frac{1}{d_E^2\,\hat\gamma}  \sum_{i,j} \int dU U_{ii} U_{jj}^* =  \frac{1}{d_E^3\,\hat\gamma}  \sum_{i,j} \delta_{ij}  =  \frac{1}{d_E^2 \,\hat\gamma}\,,
 \label{eq:uav}
\end{equation}
where ${\bf E}_\text{U}$ stands for the average over the uniform distribution of unitary matrices.\footnote{In a previous version of the paper, we obtained an estimate scaling as $1/d_E$ by diagonalizing $\text{U}$ and assuming that the eigenvalues $e^{i \theta_i}$ are drawn uniformly from the circle. This is not correct, as the Jacobian factor in passing from the integral over $U_{ij}$ to the eigenvalues introduces an eigenvalue repulsion. This repulsion produces the additional suppression in the average correlator found here.} The average value of the two-sided correlation function is smaller than the single-sided one by a factor of $1/d_E^2 = e^{-2S}$. 

We will see below that the standard deviation in the average over operators is larger than this average, scaling as $e^{-S}$, so the correct estimate for the two-sided correlator for a particular operator is smaller than the single-sided one by a factor of $1/d_E = e^{-S}$.  We conclude that these typical states cannot have a smooth wormhole description in the gravitational dual. The same conclusion was reached in \cite{Marolf:2013dba} by appealing to the eigenvalue thermalization hypothesis \cite{deutsch,srednicki}. Our approach based on random matrices gives a different, more computationally tractable perspective on the result. 

It would be interesting to study the behaviour of this two-sided correlation function for  small deformations of $\text{U}=\mathbb{I}_E$, and compare it to the changes in the length of the wormhole, which were recently studied by  Shenker \& Stanford \cite{Shenker:2013pqa,Shenker:2013yza}. We leave this for future work.

\subsection{Standard deviations}

Deviatons within the ensemble give us a measure of the departure of the correlator for a specific operator from the ensemble average considered above. We would expect these deviations to be small when the dimension of the Hilbert space is large. 

The standard deviations in our operator ensemble are
\begin{eqnarray}
  \sigma^2_{{\cal O},RR}(\text{U},t) = & {\bf E}\bigl ( | \langle \psi_\text{U} | O^\dagger_R O_R | \psi_\text{U}\rangle |^2 \bigr) -  | {\bf E}\bigl ( | \langle \psi_\text{U} | O^\dagger_R O_R | \psi_\text{U}\rangle \bigr) |^2, \label{eq:vari1} \\
  \sigma^2_{{\cal O},RL}(\text{U},t) = & {\bf E}\bigl ( | \langle \psi_\text{U} | O^\dagger_R O_L | \psi_\text{U}\rangle |^2 \bigr) -  | {\bf E}\bigl ( | \langle \psi_\text{U} | O^\dagger_R O_L | \psi_\text{U}\rangle \bigr) |^2.  \label{eq:vari2}
\end{eqnarray}
The one-sided quantity \eqref{eq:vari1} will obviously be independent of the unitary matrix $U$. It is\begin{multline}
  \sigma^2_{{\cal O},RR}(\text{U},t) = e^{-2S(E)}\,\\
  \sum_{i,j,k,l} e^{i(E_i-E_j)t}\,e^{-i(E_k-E_l)t}\,\left({\bf E}\left(M_{ji}^\star M_{ji}M_{lk}M_{lk}^\star\right) - {\bf E}\left(M_{ji}^\star M_{ji}\right){\bf E}\left(M_{lk}^\star M_{lk}\right)\right)\,.
\end{multline}
The only non-trivial contribution comes from the contractions ${\bf E}(M_{ji}^\star M_{lk}){\bf E}(M_{ji}M_{lk}^\star)$. These give
\begin{equation}
   \sigma^2_{{\cal O},RR}(\text{U},t) = \frac{e^{-2S(E)}}{\hat\gamma^2}\,.
\end{equation}
Note that the index contractions are such that the phases cancel, so the variation is time-independent. The computation of \eqref{eq:vari2} is very similar. In this case,
\begin{multline}
  \sigma^2_{{\cal O},RL}(\text{U},t) = e^{-2S(E)}\,\\
  \sum_{i,j,k,l} U_{ij}U^\star_{kl}e^{i(E_l-E_j)t}\,U^\star_{i^\prime j^\prime}U_{k^\prime l^\prime}e^{-i(E_{l^\prime}-E_{j^\prime})t}\,\left({\bf E}\left(M_{jl}^\star M_{ik}M_{j^\prime l^\prime}M_{i^\prime k^\prime}^\star\right) - {\bf E}\left(M_{jl}^\star M_{ik}\right){\bf E}\left(M_{i^\prime k^\prime}^\star M_{j^\prime l^\prime}\right)\right)\,.
\end{multline}
Its non-trivial contribution comes from the same contractions as before, and the index contraction is such that the factors of $U$ cancel out in addition to the time dependence, giving
\begin{equation}
  \sigma^2_{{\cal O},RL}(\text{U},t) = \frac{e^{-2S(E)}}{\hat\gamma^2}\,.
\end{equation}

We can compare these standard deviations to the average size of the correlators, 
\begin{equation}
\begin{aligned}
  \frac{\sigma^2_{{\cal O},RR}(\text{U},t)}{| {\bf E}\bigl ( | \langle \psi_\text{U} | O^\dagger_R O_R | \psi_\text{U}\rangle \bigr) |^2} &= \frac{e^{2S}}{|\text{tr} W(t)|^4}, \\
  \frac{\sigma^2_{{\cal O},RL}(\text{U},t)}{| {\bf E}\bigl ( | \langle \psi_\text{U} | O^\dagger_R O_L | \psi_\text{U}\rangle \bigr) |^2}&= \frac{e^{2S}}{|\text{tr}\, \text{U}W(t)|^4}.
\end{aligned}
\end{equation}
For the single sided correlators, so long as  $t\Delta \ll 1$ (where $\Delta$ is the energy spread in $\mathcal H_E$)  the standard deviation is small compared to the average, as expected. For the two-sided correlators, the answer depends on the state under consideration, i.e. the unitary matrix $U$. For the standard purification $\text{U}=\mathbb{I}$, the one sided and two sided correlators are the same size. But for typical unitary matrices $\text{tr} \text{U} \sim e^{S/2}$, so already  for $t =0$, the standard deviation is larger than the correlator. Thus, for a typical state $| \psi_{\text U} \rangle$, the overall size of the correlator for a particular operator should be estimated from the standard deviation; the typical value is thus smaller than the single-sided one by $e^{-S}$, and the value fluctuates from operator to operator, producing a smaller average value. This also seems problematic for attempts to interpret the correlations as due to a smooth semiclassical wormhole. 

\section{Operator ensemble including energy-changing transitions}
\label{can}

In the previous section, we assumed that the operator changes the energy only by a small amount. This restriction may be too strong for some operators, so in this section we will provide a model for random operators when allowing for transitions between more disparate energy states. This will be useful for considering correlations in states corresponding to the canonical ensemble, and will allow us to model the bulk time-dependence associated with quasinormal modes  of the black hole. 

When we allow transitions that change the energy,  not all states are on an equal footing and the distribution of matrix elements can change as a function of the initial and final state energies. The operator ensemble we consider to model this behaviour should then be more complicated. Following the philosophy of effective field theory, we write this matrix distribution as an expansion in the energy separation of the states involved in the transition, assuming the existence of an averaged energy $E$ around which we work. The leading terms in the distribution of matrices are then
\begin{equation}
  {\cal F}_{\text{g}}=\frac{1}{Z^M_{\text{g}}} \prod_{ij}\,dM_{ij}dM^*_{ij}\,  e^{-\gamma \bigl( \text{tr}(M M^\dagger) - \alpha_1 \text{tr}([M,H]M^\dagger)+\alpha_2 \text{tr}([H,M][M^\dagger,H])+...\bigr)},
 \label{eq:can0}
\end{equation}
where $Z^M_{\text{g}}$ ensures the normalisation condition $\int {\cal F}_{\text{g}}=1$ is satisfied. We will refer to this as the {\it energy-changing operator ensemble}.

In the spirit of effective field theory, the parameters $\alpha_i$ will be determined by matching the properties of the correlators in the ensemble to those of correlators in a bulk black hole for some particular field. We will see that the resulting values guarantee convergence of the matrix integrals in $\int {\cal F}_{\text{g}}$.  Also, a priori $|i\rangle,|j\rangle$ run over all the states in the Hilbert space, but transitions between highly separated energies are suppressed in the ensemble by  $\alpha_2$,  and the size of the energy transitions for the values determined from the black hole will be small compared to the overall energy in the thermodynamic limit.

Since we retain in ${\cal F}_\text{g}$ only terms quadratic in the matrix elements, we can rewrite \eqref{eq:can0} as
\begin{equation}
  {\cal F}_\text{g}  \propto \Pi_{ij} dM_{ij}dM^*_{ij}  e^{-\gamma \sum_{kl} \Delta_{kl} |M_{kl}|^2} ,
  \end{equation}
  with
  \begin{equation}
    \Delta_{kl} = 1 +\alpha_1 (E_k-E_l) +   \alpha_2 (E_k-E_l)^2+ \dots \equiv P(\alpha_j, E_k-E_l),
\label{eq:cancor}
\end{equation}
where in the last step we introduced a general polynomial of the energy difference $P(E_k-E_l)$ to emphasize that many of our statements can be extended to higher orders in $E_k-E_l$. For the most part we will truncate the expansion to second order, but
when specific details will not matter, we will use the arbitrary polynomial $P$. As in subsection \ref{sec:micro}, the only non-trivial connected correlation function is 
\begin{equation}
  {\bf E}\left(M^\star_{ij}M_{kl}\right) = \frac{1}{\gamma\,\Delta_{ij}}\,\delta_{ik}\delta_{jl}\,.
\end{equation}

Let us first discuss the scaling of $\gamma$ in this ensemble. As in \eqref{eq:gfix}, we will require the total ``inclusive cross-section" for transitions between any states of energy $E_i$ and $E_j$ to be finite. We would now like to do this while preserving the symmetry of the transition amplitude $M_{ij}$ in $i$ and $j$.  It is then  convenient to choose $\gamma$ as 
\begin{equation}
\gamma={\hat\gamma} \bigl( e^{S(E_i)}+e^{S(E_j)} \bigr).
\end{equation}
This entails a minor modification of \eqref{eq:can0} that moves $\gamma$ inside the trace\footnote{Strictly speaking $\hat\gamma$ should also be inside the trace. However, this will have no consequences for our conclusions.}. There is some arbitrariness in our definition of the scaling of $\gamma$, but in the effective field theory approach that we are discussing, and when we allow transitions only between close by $E_k$ and $E_l$, then other choices which keep the ``inclusive cross section'' finite amount to redefinitions of the $\alpha$s. We will use the choice above because it is computationally simple.  

The ``inclusive cross section" is then
\begin{equation}
\sum_{k\in {\cal H}_{E_j}} |\langle k | M | i_0\rangle|^2\sim {e^{S(E_j)} \over {\hat \gamma} \bigl (e^{S(E_j)}+e^{S(E_{i_0})}\bigr)\,P(\alpha_l,E_j-E_{i_0})},
\end{equation}
which is finite for all $|i_0 \rangle$ and $E_j$. We should actually require the above expression to be finite when we integrate over $E_k$, but we will see that this happens automatically because of the $\text{tr}([M,H]^2)$ terms.

Let us now determine the coupling constants $\alpha_l$ and understand their physical relevance by matching observables. As said earlier, we will choose these by matching the correlators in the ensemble to those of a particular operator. That is, we now model a given operator $\mathcal O$ as a matrix chosen at random from the ensemble \eqref{eq:can0} with ensemble parameters $\alpha_l$ chosen to reproduce some of the expected structure of the operator $\mathcal O$. Consider the one sided two point function in the canonical ensemble,
\begin{equation}
\begin{aligned}
  \frac{1}{Z(\beta)} {\bf E}\left(\text{tr}(e^{-\beta H} {\cal O}_R^\dagger(t){\cal O}_R(0))\right) &= \frac{1}{Z(\beta)} \sum_{ik} e^{-\beta E_i}\,e^{i (E_i-E_k) t}\, {\bf E}\left(M^*_{ki} M_{ki}\right) \\
 &= \frac{1}{Z(\beta)} \int dE_i dE_k \frac{e^{-\beta E_i + S(E_i)+S(E_k)+i(E_i-E_k)t}}{\hat\gamma\,\bigl (e^{S(E_i)}+e^{S(E_{k})}\bigr)P(\alpha_j,E_k-E_i)}.
\end{aligned}
\label{TmCan1}
\end{equation}
Note that this integral is convergent, as large $E_i$ are cut-off by the $-\beta E_i$ term and for large $E_k$, the numerator and denominator $e^{S(E_k)}$ cancel, leaving us with terms which are oscillating in $E_k$. Changing variables to $E_i$ and the energy difference $\Delta = E_k-E_i$, the integral becomes approximately
\begin{equation}
\int d\Delta  \frac{e^{-it \Delta}}{{\hat\gamma}(1+e^{-\beta\Delta})\,P(\alpha_l, \Delta)}.
\label{TmCan2}
\end{equation}
In the last formula we cancelled the $E_i$ integral against $Z(\beta)$ (picking up corrections which scale like $1/E$ where $E$ is the averaged energy, as we show in the appendix).\footnote{To do so we expanded $S(E_k)=S(E_i-\Delta E)$ to 1st order in $\Delta E$. This approximation is valid since the $\Delta E^2$ term in this expansion is multiplied by a quantity which is of order $1/E$ but $\Delta E$ is finite and does not scale with the total energy, so this is small. Similarly the $\alpha$'s depend on the energy in which we evalute them - ie, they hide an $E_i$ dependence. However, taking this dependence into account introduces correction terms proportional to $\partial_E \alpha$ and $\partial_E \beta(E)$. These terms again scale as $1/E$ and hence are negligible.}

For large $t$ we deform the contour of $\Delta$ to the upper half plane and pick up the leading pole of the denominator. Denoting it by $\Delta_0,\ \text{Im}(\Delta_0)>0$, the correlator decays exponentially at long times as  
\begin{equation}
e^{i\Delta_0 t}.
\end{equation}
In the bulk spacetime, the exponential decay of a particular correlator is determined by the lowest quasi-normal mode of the corresponding field; i.e., $\Delta_0$ corresponds  to the complex frequency of the lowest quasi-normal mode of the bulk black hole.  Thus, given a bulk field, we choose the coefficients $\alpha_l$ in the definition of the ensemble to reproduce the quasi normal modes of the bulk field. In the quadratic approximation we focus on, $P(\alpha_l,\Delta) = 1 + \alpha_1 \Delta + \alpha_2 \Delta^2$, and $\alpha_{1,2}$ are determined by requiring that this polynomial has a zero at the complex frequency of the lowest quasi-normal mode. If we retained higher-order terms in the polynomial, these could be determined by requiring further zeros match the complex frequencies of higher quasi-normal modes.

Additionally, $\alpha_1$ encodes the commutator $[{\cal O}^\dagger_R,{\cal O}_R]$ since
\begin{equation}
\begin{aligned}
  {\bf E}\bigl( \text{tr}(\frac{e^{-\beta H}}{Z(\beta)} [{\cal O}_R^\dagger,{\cal O}_R])\bigr) & = \frac{1}{Z(\beta)} \sum_{i,k} e^{-\beta E_i} {\bf E}\left(M^*_{ki}M_{ki} - M_{ik}M^*_{ik}\right) \\
  &= \frac{1}{Z(\beta)} \sum_{i,k} {e^{-\beta E_i}\over \gamma} \left(\frac{1}{\Delta_{ki}}-\frac{1}{\Delta_{ik}}\right) 
\end{aligned}
\end{equation} 
%For simplicity of the expressions, we will assume from now on that $\alpha_1=0$, which corresponds to neglecting the real part of the quasi-normal mode frequency. 

In the restricted operator ensemble we truncated to $E_l=E_k$ and dropped the commutator terms in the brackets. This can be realized as a limit of our more general operator ensemble by rescaling $\alpha_2 \to \infty,\,\alpha_1\to 0$ with an appropriate rescaling of ${\hat \gamma}$. Thus, the restricted operator ensemble will be a good approximation for times short compared to the lowest quasi normal frequency of the bulk field. 

\subsection{Single sided correlators}

Having determined the scaling behaviour of the parameters characterizing our operator ensemble \eqref{eq:can0}, we can now calculate ensemble averages of correlation functions. In this subsection we consider the single-sided correlators, which depend only on the reduced density matrix obtained from the state. 

We calculated the one-sided two-point function for the thermal density matrix in \eqref{TmCan1}, and fixed the parameters $\alpha_l$ by requiring that it have an exponential decay in time which corresponds qualitatively to the expected exponential decay of correlations due to the quasi-normal modes in the bulk black hole. Reproducing this exponential decay (that is, having a corresponding short time scale in the operator ensemble) was one of the primary motivations for generalizing from the restricted operator ensemble. 

We can also easily calculate the one-sided two-point function in the microcanonical ensemble of states specified by an energy $E$, $\text{tr}\left(\rho_E{\cal O}^\dagger_1(t){\cal O}_1(0)\right)$. The difference from the previous section is that we allow intermediate states at energies different from $E$. This gives
\begin{equation}
\int dE_k { e^{S(E_k)+i(E_i-E_k)t} \over {\hat \gamma} \bigl(e^{S(E_i)}+e^{S(E_k)}\bigr)  (1+\alpha_1 (E_k-E_i) +\alpha_2(E_k-E_i)^2)},
\label{TmMcan1}
\end{equation}
which is the same integral as we had in \eqref{TmCan1} (after switching to $\Delta E=E_i-E_k$ and expanding for small $\Delta E$). The fact that we get the same result from the canonical and microcanonical ensemble of states (up to $1/E$ corrections) is what we expect from standard thermodynamic considerations. 

We can easily extend the calculation of the one-sided two-point function to an arbitrary pure state in ${\cal H}_L\otimes {\cal H}_R$
\begin{equation}
  |\psi_c\rangle =\sum_{i,j\in{\cal H}} c_{ij} |i,j\rangle\,,
\label{entstat}
\end{equation}
where $\sum_{i,j\in {\cal H}} |c_{ij}|^2=1$. This includes as particular cases: 
\begin{itemize}
\item Microcanonical entangled state: $c_{ij} = e^{-S(E)/2} \delta_{ij}$ for $|i \rangle, |j \rangle \in \mathcal H_E$, zero otherwise
\item Thermofield double: $c_{ij} = \delta_{ij} e^{- \beta E_i/2}$
\item Unitary twisting of the microcanonical entangled state (which reduces to the microcanonical ensemble on tracing over one side): $c_{ij} = e^{-S(E)/2} U_{ij}$ for $|i \rangle, |j \rangle \in \mathcal H_E$, zero otherwise
\item Unitary twisting of the thermofield double: $c_{ij} = U_{ij} e^{-\beta E_i/2}$ 
\item Pure state: $c_{ij} = V^*_i V_j$. 
\end{itemize}

One sided correlators only depend on the reduced density matrix
\begin{equation}
  \rho_c= \text{tr}_{{\cal H}_L} |\psi_c\rangle \langle\psi_c | = \sum_{ij} (c^\dagger c)_{ji} |i\rangle \langle j|\,.
\end{equation}
The one sided two-point function equals
\begin{equation}
  \text{tr}\left(\rho_c\,{\cal O}_R^\dagger(t){\cal O}_R(0)\right) = \sum_{i,j,m}  (c^\dagger c)_{ji}\,e^{it(E_j-E_m)}\,M_{mi}M^\star_{mj}\,,
\end{equation}
 thus its operator ensemble average in the energy-changing ensemble equals
\begin{equation}
  {\bf E}\left(\text{tr}\left(\rho_c\,{\cal O}_R^\dagger(t){\cal O}_R(0)\right)\right) = \sum_{i,m} (c^\dagger c)_{ii}\,\frac{e^{it(E_i-E_m)}}{\gamma\,\Delta_{mi}}\,.
\label{eq:gen1sided}
\end{equation}

In this case the correlator is
\begin{equation}
\int dE_k\,\sum_i (c^\dagger c)_{ii} { e^{S(E_k) +i (E_i-E_k)t} \over {\hat\gamma}  (e^{S(E_i)}+e^{S(E_k)}) (1 + \alpha_1(E_k-E_i)+\alpha_2 (E_k-E_i)^2 ) }
\label{genc}
\end{equation}
Recall that our normalization condition is $\sum_i (cc^\dagger)_{ii}=1$. The particular cases considered above are sharply peaked around some energy $E$ (at least when the ensembles are stable) so in all these cases the correlator is qualitatively the same. In particular this is true for a pure state $c_{ij} = V^*_i V_j$. This is just a consequence of the fact that we can divide our states into very fine energy slices, and then the ensemble is invariant under arbitrary unitary transformations in each of the slices. Thus the single-sided probes are not sensitive to any detailed information about the state. 

\subsubsection{Standard deviations}  
\label{sec:stan2}

To study how reliable the ensemble averages \eqref{eq:gen1sided} are, we compute the variances of single sided two-point correlators in our operator ensemble, 
\begin{equation}
  \sigma_{cRR}^2 = {\bf E}\left(\left| \text{tr}\left(\rho_c {\cal O}^\dagger_R(t) {\cal O}_R(0)\right) \right|^2 \right) - \left|{\bf E}\left(\text{tr}\left(\rho_c {\cal O}^\dagger_R(t){\cal O}_R(0)\right)\right)\right|^2\,.
\label{eq:variance2}
\end{equation}
The first term is 
\begin{equation}
  {\bf E}\left(\left[ \text{tr}\left(\rho_c {\cal O}^\dagger_R(t) {\cal O}_R(0)\right) \right]^2 \right) = \sum_{i,j\dots m,n} (c^\dagger c)_{ij}(c^\dagger c)^\star_{kl}\,e^{it(E_j-E_m)}e^{-it(E_l-E_n)}\langle M_{mi}M^\star_{mj}M^\star_{nk}M_{nl}\rangle\,.
\end{equation}
The only contribution to the difference \eqref{eq:variance2} comes from contractions in which the operators ${\cal O}_R$ are contracted between traces.
Thus, we are only left with the contraction $\langle M_{mi}M^\star_{nk}\rangle \langle M_{mj}M^\star_{nl}\rangle$. Thus the variance is 
\begin{equation}
  \sigma_{cRR}^2 = \sum_{i,j,m} \frac{|(c^\dagger c)_{ij}|^2}{\gamma^2}\frac{1}{\Delta_{mi}\Delta_{mj}} \,=\, \sum_{i,j} {|(c^\dagger c)_{ij}|^2}\int dE \frac{e^{S(E)}}{\gamma^2P(E-E_i)P(E-E_j)}\,,
\end{equation}
where we took the continuum limit in $E_m=E$ in the last step.

Plugging in the expressions for $c_{ij}$ in the canonical, microcanonical and pure state we obtain the following scaling behaviour for $\sigma_{cRR}^2$: $\text{Canonical},\ \text{Microcanonical}\,\sim e^{-2S(E)},\  \text{Pure}$ $\text{state}\,\sim  e^{-S(E)}$.

\subsection{Two sided correlators}

In this section we compute the two-sided correlation function in the general pure state $|\psi_c \rangle$ defined in \eqref{entstat}. Applying the map of operators from ${\cal H}_R$ to ${\cal H}_L$ discussed in section \ref{sec:map}, the two sided two-point function in an arbitrary pure state $|\psi_c\rangle$ is
\begin{equation}
   \langle \psi_c | {\cal O}_R^\dagger(t) {\cal O}_L(0) |\psi_c \rangle = \sum_{i,j,k,l} c_{ij}c^\star_{kl}\,e^{it(E_l-E_j)}\,M_{ik}M^\star_{jl}\,.
 \end{equation}
Thus, its operator ensemble average is
\begin{equation}
  {\bf E}\bigl( \langle \psi_c | {\cal O}_R^\dagger(t) {\cal O}_L(0) |\psi_c \rangle  \bigr) = \sum_{i,k} c_{ii}c^{*}_{kk} \frac{e^{i(E_{k}-E_{i})t}}{\gamma\,\Delta_{ik}}\,.
\label{eq:2sidedgen} 
\end{equation}
This should be compared with the single sided correlator for the same $|\psi_c\rangle$, in equation \eqref{eq:gen1sided}.

For the special cases corresponding to the canonical and microcanonical ensemble where $c_{ij} \propto \delta_{ij}$, we can see that the correlators have the same form as the single-sided one. 
For the unitary transformed state with  $c_{ij}=U_{ij}e^{-\beta E_j/2}$, where $U$ is an arbitrary unitary matrix, as in the restricted operator ensemble, the two sided correlator will depend on $U$. We define the "coherence density" as
\begin{equation}
F(E)=e^{-S(E)} \sum_{i\in {\cal H}} U_{ii},
\end{equation}
then the two sided correlator is
\begin{equation}
{\bf E}\bigl( \langle \psi_c | {\cal O}_R^\dagger(t) {\cal O}_L(0) |\psi_c \rangle  \bigr) = {1\over Z} \int dE_i dE_k F(E_i)F^*(E_k) {e^{-\beta E_i/2-\beta E_k/2+S(E_i)+S(E_k)}\over e^{S(E_i)}+e^{S(E_k)}} {e^{i(E_k-E_i)t}\over {\hat \gamma} \Delta_{ik}(E_i,E_k)}
\end{equation}
For $U_{ij} = \delta_{ij}$, $F$ is of order one, and this two-sided correlator is similar to the one-sided correlator. For generic $U_{ij}$, $F$ is exponentially suppressed, and as in our previous discussion in the microcanonical operator ensemble, the two-sided correlator is exponentially suppressed relative to the single-sided correlator, so we conclude that the states do not have a smooth wormhole description in the dual. 

Again, it would be interesting to compare cases where $F(E)$ is order one to the discussion of semi-classical throats of greater length in \cite{Shenker:2013pqa}. For the canonical ensemble, we could take $c_{ij}=e^{-\beta E_i/4-\beta E_j/4}V_{ij}$ where $V$ is close to the identity, and mixes only states very close in energy, say less than $\delta$. In this case $F(E)=e^{-S(E)+{\cal O}(\beta\delta)}\sum_{i\in {\cal H}_E}V_{ii}=1+{\cal O}(\beta\delta)$. 

\subsubsection{Standard deviations}

To study how reliable the ensemble averages \eqref{eq:2sidedgen}  are, we compute the standard deviations of two sided two-point correlators in our operator ensemble,
\begin{equation}
  \sigma_{cRL}^2 = {\bf E}\left(\left| \langle\psi_c |{\cal O}^\dagger_R(t) {\cal O}_L(0)|\psi_c \rangle\right|^2 \right) - \left|{\bf E}\left(\langle\psi_c |{\cal O}^\dagger_R(t){\cal O}_L(0)|\psi_c \rangle\right)\right|^2\,.
\label{eq:variance3}
\end{equation}
The calculation proceeds as in section \ref{sec:stan2}. In particular, the origin of the contractions giving rise to a non-trivial contribution is entirely the same. The variance equals
\begin{equation}
  \sigma_{cRL}^2 = \sum_{i,j,k,l} \frac{|c_{ij}|^2|c_{kl}|^2}{\gamma^2}\,\frac{1}{\Delta_{ik}\Delta_{jl}}\,.
\end{equation}
For the $c_{ij}$ corresponding to both the standard ensemble states and their generalized versions with a unitary matrix, this scales as $e^{-2S}$, as in the discussion in the restricted operator ensemble. For the standard  microcanonical and canonical states this then implies the standard deviation is  small compared to the average value, while for the general states involving $U_{ij}$, it is of the same order as the average value.

Thus the results for two-sided correlators in this more general energy-changing operator ensemble are qualitatively the same as in the simpler restricted ensemble. 

\subsection{Relation to the eigenvalue thermalization hypothesis}

We have introduced these random matrix ensembles as a way to model the behaviour of operator correlation functions in a complicated thermal system (namely the black hole states). We would like to compare this to a previous description of such correlators, the eigenvalue thermalization hypothesis (ETH) \cite{deutsch,srednicki}. The ETH assumes\footnote{We note that the separation of the operator matrix elements into a smooth $f$ and $M^L$ seems somewhat artificial, since both of them are really matrices that depend of $\alpha$ and $\beta$, and hence it's not clear where to draw the line between them.}
\begin{eqnarray}
& \langle k | \mathcal O_L | i \rangle = {\cal A}({\bar E} )\delta_{ki} + e^{-S({\bar E})/2} f^A(E_\alpha,E_\beta) M^L_{\alpha\beta} \\
& \langle k | \mathcal O_R | i \rangle = {\cal B}({\bar E} )\delta_{ki} + e^{-S({\bar E})/2} f^B(E_\alpha,E_\beta) M^R_{\alpha\beta} \\
& M^L_{ki}M^R_{mn} = \delta_{km}\delta_{in}\sigma^{LR}(E_\alpha,E_\beta)\ +\text{erratic}
\end{eqnarray}
In our analysis, we have assumed that the one point functions ${\cal A}$ and ${\cal B}$ are approximately vanishing; this is typical for non-conserved quantities in the thermodynamic limit, but our model could easily be extended to include such terms. The main point of comparison is then the second part of the ETH matrix elements. 

Our previous microcanonical operator ensemble can be compared to the ETH matrix elements with  $E_\alpha=E_\beta$. These are matrices whose only non-trivial connected Green's function is a two point function + erratic. This corresponds precisely to the behaviour of our microcanonical ensemble in \eqref{eq:2ptmc}; the fluctuations of individual elements of the ensemble about the average will produce the erratic term. (By erratic one means a random noise which averages to zero.) So our restricted operator ensemble is one example of an ensemble that reproduces the behaviour assumed in ETH, for operators which make transitions between nearby energies. 

In the energy-changing ensembles, our correlators have again the structure assumed in ETH, but we have introduced a different description of the dependence on the energy differences. In ETH, this is characterized by the functions $f^A$, $f^B$ and $\sigma^{LR}$, which are assumed to be smooth functions of $E_\alpha, E_\beta$, but are otherwise arbitrary. Our model goes a step further, by providing a simple  ``effective action" description of this dependence for energies $E_\alpha, E_\beta$ near some fixed energy. This effective action approach allows us to fix the energy dependence to reproduce aspects of the black hole behaviour like the quasinormal mode. This provides a nice way to build more specific models of the operator matrix elements; our approach could easily be extended to include features of interest for other specific applications.

\section{Structured operators and the confined phase}
\label{lowT}

In this section, we first note that there are cases in which a thermal density matrix does not correspond to an Einstein-Rosen wormhole, but the two sided correlators are still large. This is  a simple General Relativity statement, which has been somewhat neglected in the discussion of connection between entanglement and geometry. We then conjecture that this may be related to the fact that the probing gravity operators are not random on the thermal states, but rather they are ``structured" in the sense of section 2. In general we want to speculate that the random nature of low-energy gravity probes is essential for understanding how entanglement is related to geometry.

It is well-known that if we consider a field theory on the sphere, there is a Hawking-Page phase transition at a finite temperature (of order the radius of the sphere) where the thermal AdS and black hole saddle-points exchange dominance. Thus, for a thermofield double state at sufficiently low temperature, the dual bulk description does not involve a large wormhole; rather, it is two copies of thermal AdS, with the thermal graviton gas on the two spaces in an entangled state. Nonetheless, the two-sided two-point functions remain of the same order as the one-sided ones. For the field theory, this is automatic: since we are in a thermal state, the one- and two-sided correlators are related by analytic continuation as in section \ref{sec:map}. In the bulk, these large two-sided correlators can be understood through the existence of a connection between the two boundaries in the Euclidean section. (So we can still interpret the two-sided correlator in terms of a geodesic length, $\langle \mathcal O_L \mathcal O_R \rangle \sim e^{-\Delta \ell}$, where $\ell$ is now the length of a geodesic in the Euclidean instanton linking the two boundaries.)  But there is no Lorentzian wormhole linking the two AdS spaces.  From the Lorentzian point of view, these correlators are large because of the entanglement between the bulk modes in the two thermal AdS spaces.

The von Neumann entropy of the reduced density matrix on one side below this transition is order one, so one might think that the absence of a smooth wormhole in this case is associated with the small amount of entanglement between the two sides. But if we work in the microcanonical ensemble, we can increase the energy to reach regimes where the entropy is large where the disconnected saddle is still the dominant bulk solution.\footnote{In the microcanonical ensemble, the thermal gas of gravitons is the dominant phase up to a Hagedorn transition, at energies set by the 't Hooft coupling in the field theory; thus the entropy will grow up to some power of the 't Hooft coupling \cite{Banks:1998dd}.} 

In our analysis, this phase is qualitatively different from the black hole phase because the gravity mode operators acting on thermal AdS will not behave like random operators; probing a thermal gas in AdS, these operators are sensitive to the differences between different states. They are sparse operators which are related in a specific way to the states that make up the ensemble, i.e, they are ``structured" with the respect to the states. Thus, it is possible that what makes the difference between the geometric connection in the black hole case and its absence here is that in the former we had large correlations for random operators, whereas here we have them for some operators which are structured with respect to the states. 

We now present a simple model to make the point that gravity modes will be structured operators in this phase. Consider a Hilbert space made out of many harmonic oscilators with frequencies $w_1,....w_k$, with total energy $H=\Sigma n_i w_i$. In the AdS/CFT picture, these are the energies of the single particle states in the bulk. We will denote the states by $|{\vec n}\rangle$ where ${\vec n}$ is a vector of integers of length $k$, which tells us the particle number for each energy level $w_i$. When we need to distinguish a specific particle - particle $l$ - from the rest we will denote the state as $| {\vec n},n_l\rangle$. The operators that we use to probe the geometry are then the corresponding raising or lowering operators.  That is,  we have at our disposal operators $T_i,\ i=1..k$, and our model for their action is \begin{equation}
<{\vec n},n_i|T^i|{\vec m},m_i>=\delta_{{\vec n},{\vec m}}\bigl(\delta_{n_i+1,m_i}+\delta_{n_i,m_i+1}\bigr)
\label{strcOp}
\end{equation} 
where the i'th index is dropped from ${\vec n}$ and ${\vec m}$. These operators are clearly not generating random transitions between the energy eigenstates; instead, the operator matrix elements are sparse matrices. 

Consider now the state 
\begin{equation}
{1\over \sqrt{Z}} \sum_{{\vec n}} e^{-\beta E_{\vec n}/2} |{\vec n},{\vec n}\rangle.
\end{equation}
This could be thought of as either 1) a model of the canonical ensemble in the full CFT for temperatures below the Hawking-Page transition, or  2) a canonical description of the gas of gravity modes when pushed to  $T \gg 1$, in the regime where it is locally stable. We can compare the single and two sided correlators of the structured operators $T_i$ in this state. A simple calculation gives 
\begin{equation}
\langle T^{(1)}_1(t) T^{(1)}_1(0) \rangle =  e^{-iw_1t}+e^{iw_1 t -\beta w_1},
\label{strsngl}
\end{equation}
while 
\begin{equation}
\langle T^{(2)}_1(t) T^{(1)}_1(0)\rangle = \bigr( e^{-iw_1t}+e^{iw_1t}\bigr)  e^{-\beta w_1/2}  .
\label{strdbl}
\end{equation}
The interpretation is simplest for the case that $w_1\beta \gg 1$, i.e, a probe which is heavy compared to the temperature. In this case the first term in \eqref{strsngl} is just the particle freely traversing the thermal space. The result in \eqref{strdbl} is suppressed by a factor of $e^{-\beta w_1/2}$.  This reproduces the suppression one would expect from propagation throughout the Euclidean instanton, where the disconnected initial time surfaces correspond to the constant time slices at $\tau =0 , \beta/2$.

\section{Black holes and random operators}
\label{rand}

We conclude with a more speculative section. The main conjecture in this paper is that low energy gravity modes are well approximated by random operators on any state which appears as a black hole for an outside observer, and that using this approximation we can find a simple criteria for when an Einstein-Rosen throat is semi-classical. Since most discussions of the information paradox \cite{infoparadox} or firewalls \cite{Almheiri:2012rt} are phrased in terms of such probes, it will be interesting to explore the consequences of their conjectured randomness in this regard.  We briefly discuss some issues in this direction, which are the robust analytic structure of such operators and the problem of ``seeing behind the horizon", and the possible emergence of a new perturbation theory for random operators, with some speculations on instances where this perturbation theory exhibits unusual behaviour for probes which cross close to the horizon.

\subsection{The analytic structure of random operators in pure states}

We modelled supergravity probes as random because they are insensitive to the details of the black hole state. One might think this would make it difficult to get any sensible coherent behaviour behind the horizon. However, precisely the opposite is true. Since the correlations don't depend on the state, we get an effective coarse-graining over the possible microstates of the black hole. As a result, the analytic structure of random operators is more robust than for more structured operators, allowing us to carry out the continuation to the ``other side" for a large class of ensembles or states on $\mathcal H_R$. 

To see this recall equations \eqref{TmCan1}, \eqref{TmCan2} and \eqref{TmMcan1}  above, which gave the same single sided two point function for any generic single sided state or density matrix (up to $1/N^2$ corrections),\footnote{By generic state we mean a state that we do not arrange using the spectral data of ${\cal O}_R$. For example we do not allow ourselves to find the eigenvectors of ${\cal O}_R$ and then choose one of them as our state. } so the single sided correlation functions do not depend on the particular state we consider, they all mimic the thermal density matrix. This means that in any such state or density matrix we can continue to the second copy in exactly the same way. This result relies heavily on the random matrix aspect of the probes, and it seems unlikely to us to hold without some assumption on the spectral properties of the probe operators. 

Thus, if we are in any of these density matrices, the value of the single-sided two point function is as in the thermal ensemble. Just as for the thermal ensemble, we can then define the action of operators on a second copy of the Hilbert space $\mathcal H_L$ as in section \ref{sec:map}. The small differences in the single-sided correlations will grow exponentially under this continuation, but it was noted in \cite{Balasubramanian:2007qv} that for a typical pure state these differences are small enough that they are still exponentially suppressed after analytic continuation to $t' = t + i \beta/2$. 

This result may seem to be in tension with our previous result that the two-sided correlators in the typical state on $\mathcal H_L \otimes \mathcal H_R$ are very different from the one-sided correlators. The point here is that we are considering the analytic continuation of the one-sided correlators, based on the fact that these do not contain information on the actual state or density matrix considered. These thus do not correspond to the actual two-sided correlators in a given typical state on $\mathcal H_L \otimes \mathcal H_R$.

\subsubsection{Relation to the Papadodimas-Raju construction}

Our conclusion that for random operators, the operators on the ``other side" can be defined independent of the choice of state or ensemble is rather similar to the premise underlying the work of Papadodimas and Raju \cite{raju}, that mirror operators can be defined for any pure state of the black hole. Our discussion of random operators provide a different perspective on when such an analytic continuation should be possible, which can be contrasted to Papadodimas and Raju's discussion in terms of large-N factorization and generalized free fields.

One of the differences is that the construction of the mirror operators in \cite{raju} was highly state-dependent, whereas our basic point was that it was the state-independence of the correlators which makes the analytic continuation robust. Our goals are different: we have discussed the analytic continuation of the operator correlation functions, whereas \cite{raju} aimed to explicitly construct an operator acting on the microscopic Hilbert space whose correlation functions reproduced these analytically continued correlators. 

It would be interesting to compare the two approaches further, and in particular whether the two approaches converge, up to small corrections, if one makes further spectral assumptions about the operators within the approach of \cite{raju}. In any case, we believe that both can be a good starting point for exploring perturbation theory in the black hole background, and deviations from it.

\subsection{Emergence of perturbation theory and modified horizon dynamics}
\label{pert}

In the standard description of low energy gravity probes, one quantizes bulk fields in the eternal black hole background and interprets their quantum state in terms of creation and annihilation operators (at least perturbatively and in the absence of backreaction). This is certainly the case in Hawking's original calculation \cite{hawking} and it is also the point made in the reconstruction of bulk operators from CFT ones in \cite{kabat,raju}. This picture becomes more involved when trying to embed this perturbative Hilbert space structure in the dual field theory Hilbert space \cite{raju,Verlinde:2013qya}.

In this standard approach, there is a well defined bulk perturbative expansion in terms of Feynman diagrams. In this subsection we ask whether there is a corresponding notion of perturbation theory in our random matrix description. 

The relation of the gaussian ensemble to free fields was discussed  in section 3.1. The gaussian matrix models gave ensemble average two-point functions
\begin{equation}
  {\bf E}\left(M_{ij}M^\star_{kl}\right) = \frac{\delta_{ik}\delta_{jl}}{\gamma\,\Delta_{ij}}\, ,
\label{eq:recap}
\end{equation}
so insertions of ${\cal O}$ are pairwise contracted, giving rise to a Wick-like expansion of correlators (after summing over the indices). 

We can speculate on how this will be extended to an interacting picture. An interaction term in the bulk, for example an $n$-particle vertex, means that we can tie $n$ propagators together. In the eikonal approximation, it means that $n$ insertions of ${\cal O}$ (or it's conjugate) can be contracted. In the matrix model, we could model this  by including higher order polynomials in our ensemble distributions \eqref{eq:microens} and \eqref{eq:can0}.  Expanding the exponents of these polynomials will give rise to perturbative insertions of these vertices in our correlator ensemble averages ${\bf E}\left(M_{ij}\dots M^\star_{kl}\right)$, which would match with the spacetime contraction. All of this, of course, applies to interactions close to or inside the black hole, within the part of the bulk correlation functions that our random matrix is supposed to model. 

This approach would correspond to diagonalizing the full Hamiltonian in the interacting theory. An alternative approach would be to write a gaussian matrix ensemble based on diagonalizing the free Hamiltonian, and write the full Hamiltonian as  $H=H_0+P({\cal O},{\cal O}^\dagger)$, for some polynomial  $P$ in the operator ${\cal O}$ and its conjugate. The interactions would then enter by including  $P$ perturbatively in the time evolution. 

But before developing perturbation theory, we can see that even at the "free field" level, the random operator Wick-like scheme has peculiar features. The reason for this is that the random matrix contraction applies to operator matrix elements rather than to the operators themselves. Consider a four-point function of the form $\langle \psi_\beta | {\cal O}_R^\dagger {\cal O}_R^\dagger {\cal O}_R {\cal O}_R |\psi_\beta \rangle$. In standard bulk perturbation theory, we would contract all possible pairings of ${\cal O}_R$ and ${\cal O}_R^\dagger$. Each such contraction would contribute equally to the full bulk correlation function, up to some time dependence, i.e., none of them would be suppressed with the dimension of ${\cal H}_E$. We want to show this is not the case within the perturbative expansion using random matrices.

In the random matrix formalism, the objects we compute are matrix element ensemble averages such as
\begin{equation}
C_1={\bf E} \left(M_2^\dagger M_1^\dagger M_2 M_1\right)\,,\quad \quad C_2={\bf E} \left(M_2^\dagger M_2 M^\dagger_1 M_1\right)\,,
\end{equation}
where we have put subscripts $1$ and $2$ on the matrices $M$ in the order of their appearance in the correlator;  all M's refer to matrix elements of the same operator. The $C_2$ correlator allows contractions $M_1-M_1^\dagger,\ M_2-M_2^\dagger$, which gives 
\begin{equation}
 {\bf E}\left(M^*_{k_1i}M_{k_1k_2}\right){\bf E}\left(M^*_{k_3k_2}M_{k_3i}\right) = \delta_{k_1k_1}\delta_{ik_2}\delta_{k_3k_3}\delta_{k_2i} = d_E^3\,.
\end{equation}
and $M_1-M_2^\dagger,\ M_1^\dagger-M_2$, which similarly gives
\begin{equation}
 {\bf E}\left(M^*_{k_1i}M_{k_3 i}\right){\bf E}\left(M^*_{k_3k_2}M_{k_1k_2}\right) = \delta_{k_1k_3}\delta_{ii}\delta_{k_1k_3}\delta_{k_2k_2} = d_E^3\,.
\end{equation}
On the other hand, for the $C_1$ correlator, while the contraction $M_1-M_2^\dagger,\ M_1^\dagger-M_2$ gives
\begin{equation}
 {\bf E}\left(M^*_{k_1i}M_{k_3i}\right){\bf E}\left(M^*_{k_2k_1}M_{k_2k_3}\right) = \delta_{k_1k_3}\delta_{ii}\delta_{k_2k_2}\delta_{k_1k_3} = d_E^3\,
\end{equation}
with the same scaling as before, the contraction $M_1-M_1^\dagger,\ M_2-M_2^\dagger$ involves 
\begin{equation}
  {\bf E}\left(M^*_{k_1i}M_{k_2k_3}\right){\bf E}\left(M^*_{k_2k_1}M_{k_3i}\right)= \delta_{k_1k_2}\delta_{ik_3}\delta_{k_2k_3}\delta_{k_1i}  = d_E\,,
\end{equation}  
and is suppressed by a factor of $e^{-2S}$. This can be understood by standard planarity arguments: the last combination involves a non-planar contraction of the matrix indices, because our Wick-like contractions in $C_1$ cross over in this case. 

From the bulk point of view, the correlators we are averaging describe the insertion and extraction of two particles. In the eikonal approximation, we can view each contraction ${\bf E}\left(M_{ij}M^\star_{kl}\right)$ as approximating a spacetime geodesic from the insertion point of ${\cal O}_R$ to the insertion point of ${\cal O}_R^\dagger$. Thus, in the bulk the distinction between the suppressed term and the other terms is that the associated bulk geodesics are crossing in the near-horizon region\footnote{We refer to the near horizon region because our random matrix description is supposed to capture the behaviour of our low energy gravity probes close to the horizon, as explained in section \ref{motiv}.} in the suppressed term (one comes out of in between the times when the other went in and came out), while the other terms involve non-crossing bulk geodesics. 

Thus, our random operator model seems to say that not all combinations of geodesics from entry to exit points contribute equally. This may represent a potentially serious  challenge for our random matrix ensemble description. However, there are known to be subtleties in determining which spacetime geodesics actually contribute to correlations and which do not \cite{Kraus:2002iv}. Alternatively, it may be a signal that our random matrix description is capturing some back-reaction effect along the lines of \cite{'tHooft:1996tq}.

The discussion above is very preliminary in several ways. First, it just involves the rules of free field theory, and no local interactions were included. Second, there is no notion of radial direction (or any bulk dimension for that matter) in our formalism, which is known to be a difficult problem in AdS/CFT. Finally, to make the range of validity of our approach more precise we need to describe more accurately the transition from the structured part of the operator to its unstructured part using the spectral properties of the underlying OPEs in the exact dual CFT description. We leave all this for future work.

\section*{Acknowledgements}
We would like to thank Ofer Aharony, Jos\'e Barb\'on, Nima Lashkari, Amir Zait and Ofer Zeitouni for interesting discussions. SFR and JS would like to thank the organisers of the COST-WIS wokshop on {\it Black holes and quantum information} for hospitality during the last stages of this project. The work of JS was partially supported by the Engineering and Physical Sciences Research Council (EPSRC) [grant number EP/G007985/1] and the Science and Technology Facilities Council (STFC) [grant number ST/J000329/1]. The work of SFR is supported in part by STFC. The work of MB is supported by an ISF center of excellence grant, GIF and the Minerva foundation.   VB is partly supported by DOE grant DE-FG02-05ER-41367.

%\newpage 

\appendix

\section{Corrections to correlators}
\label{:apB}

We have emphasized before that single sided correlators are the same for any state or ensemble, which is peaked around a given energy, in particular in the microcanonical, canonical ensembles and some of the pure states. The discussion in section 6, on ``seeing behind the horizon", hinged on this assertion. In this appendix we will quantify the differences between these expressions, i.e., examine the size of of the corrections when going from \eqref{TmCan1} and the analogue formula for the microcanonical ensemble, which is
\begin{equation}
\int dE_k {e^{S(E_k) + i(E_i-E_k)t} \over {\hat \gamma} ( e^{S(E_i)}+e^{S(E_k)} )P(\alpha,E_k-E_i) }
\label{eq:apndmc}
\end{equation}
to \eqref{TmCan2}. 

To go from \eqref{eq:apndmc} to \eqref{TmCan2} we just expand $S(E_i)=S(E_k)-\beta(E_K) \Delta+O( \Delta^2/extensive)$ where $\Delta=E_k-E_i$. We then obtain expression \eqref{TmCan2}, up to the $O(1/extensive)$ corrections, where by $1/extensive$ correction we mean corrections that scale like $1/energy$ or $1/entropy$. 
To go from the canonical ensemble expression \eqref{TmCan1} to \eqref{TmCan2} we carry out the same expansion and change the integration variables to $E_i$ and $\Delta$. This part gives rise to $1/extensive$ corrections as before. The integral over $E_i$ localizes around the average energy in the ensemble ${\bar E}$, up to $E_i - \bar{E} \sim \sqrt{extensive}$. 
We are then justified to expand $\alpha(E_i)$ around ${\bar E}$ as $\alpha(\bar E) + (1/extensive) (E_i-{\bar E})+..$ which gives rise to additional $1/extensive$ corrections. Hence, the canonical and microcanonical correlators are the same up to $1/extensive$ corrections. 
In addition, there could be $1/N^2$ corrections to these formulas, which could be larger then the $1/extensive$ corrections identified in this appendix.

\end{document}